\documentclass[eqsecnum,noshowpacs,showkeys,nofootinbib,aps]{revtex4}
\usepackage{epsfig}

\renewcommand{\theequation}{\arabic{equation}}
\newcommand\beq{\begin{equation}}
\newcommand\eeq{\end{equation}}
\newcommand\bea{\begin{eqnarray}}
\newcommand\eea{\end{eqnarray}}
\newcommand\nn{\nonumber}

\newcommand\pr{\prime}

\begin{document}
\title{Dirac quantization and baryon intrinsic frequencies in hypersphere soliton model}
\author{Soon-Tae Hong}
\email{galaxy.mass@gmail.com}
\affiliation{Center for Quantum Spacetime and Department of Physics, Sogang University, Seoul 04107, Korea}
\date{\today}
\begin{abstract}
Quantizing a soliton on a hypersphere, we obtain the first class Hamiltonian, and evaluate the baryon 
physical quantities which are in good agreement with the corresponding experimental data. In particular, we find that the predicted value for axial coupling constant is 
comparable to its experimental value. 
The prediction for delta baryon mass 
possessing the Weyl ordering correction obtained in the first class Dirac quantization is improved comparing with that in the second 
class canonical quantization performed on the hypersphere. Making use of the same input parameters associated with the baryon masses, 
we also investigate the hypersphere soliton and standard Skyrmion models to compare the corresponding predictions for 
the physical quantities effectively. Next, we evaluate the intrinsic frequencies of the pulsating baryons. We thus find that the intrinsic pulsating frequency of more massive particle is greater than that of the less massive one. We explicitly evaluate the intrinsic frequencies $\omega_{N}=0.87\times 10^{23}~{\rm sec}^{-1}$ and 
$\omega_{\Delta}=1.74\times 10^{23}~{\rm sec}^{-1}$ of the nucleon and delta baryon, respectively, 
to yield the identity $\omega_{\Delta}=2\omega_{N}$.
\end{abstract}

\maketitle

\section{Introduction}
\setcounter{equation}{0}
\renewcommand{\theequation}{\arabic{section}.\arabic{equation}}

It is well known that, the Dirac Hamiltonian scheme has been developed~\cite{di,hong15}, to convert the second class constraints 
into the first class ones. 
Moreover, exploiting the Dirac quantization in the first class formalism, there have been attempts to quantize the 
constrained systems~\cite{hong15}. The standard Skyrmion model~\cite{skyrme61,anw83,hong98plb,hong15} which is the constrained 
Hamiltonian system, has been exploited~\cite{anw83} to investigate baryon kinematics and predict its physical properties. 
In this model, the three dimensional physical space is assumed to be topologically compactified to $S^{3}$ and the spatial infinity is located on the north-pole of $S^{3}$. Next the intrinsic frequencies of baryons have been naively studied in the standard Skyrmion model~\cite{hong15un}.

On the other hand, the hypersphere soliton model (HSM)~\cite{man1,hong98plb} has been proposed to obtain a topological lower bound on soliton energy and a set of equations of motion via the canonical quantization in the second class formalism~\cite{man1}. Later the baryon physical properties, such as baryon masses, charge radii and magnetic moments, have been newly predicted using the canonical quantization in the HSM, to suggest that a realistic hadron physics 
can be delineated in terms of this phenomenological soliton~\cite{hong98plb}.

In 1962 Dirac proposed~\cite{di62} that the electron should be considered as a charged conducting surface and its 
shape and size should pulsate. Here the surface tension of the electron was supposed to prevent the electron 
from flying apart under the repulsive forces of the charge. Motivated by his idea, we will investigate pulsating baryons in the first class formalism 
in the HSM~\cite{man1,hong98plb}, to evaluate the intrinsic frequencies of 
the baryons, baryon masses with the Weyl ordering correction (WOC) and axial coupling constant. 

In this paper, we will explicitly evaluate the intrinsic frequencies of the 
pulsating baryons.  To do this, we will use 
the baryon Hamiltonian spectrum which can be obtained through the Dirac quantization. Moreover in the HSM we will 
systematically generalize the predictions~\cite{hong98plb} for baryon physical quantities obtained in the second class formalism, to 
those in the first class one. 

In Sec. II, we will recapitulate the baryon formalism in the HSM in the second class canonical quantization scheme. 
In Sec. III, we will newly predict the corrected baryon masses, and the axial coupling constant in the HSM in 
the first class Dirac quantization. In Sec. IV, we will explicitly evaluate the intrinsic frequencies of the 
pulsating baryons such as nucleon and delta baryon. Sec. V includes conclusions.

\section{Set up of baryon formalism in HSM}
\setcounter{equation}{0}
\renewcommand{\theequation}{\arabic{section}.\arabic{equation}}

In this section, we recapitulate the baryon formulation in the HSM~\cite{man1,hong98plb}, by introducing 
the Skyrmion Lagrangian density
\beq
{\cal L}=\frac{f_{\pi}^{2}}{4}{\rm tr}(\partial_{\mu}U^{\dagger}
         \partial^{\mu}U)+\frac{1}{32e^{2}}{\rm tr}[U^{\dagger}\partial_{\mu}U,
         U^{\dagger}\partial_{\nu}U]^{2},
\label{lagtot}
\eeq
where $U$ is an SU(2) chiral field. Note that the HSM includes two parameters, 
namely a pion decay constant $f_{\pi}$ 
and a dimensionless Skyrme parameter $e$. We may fix the parameters by exploiting mesonic processes such as the decay of the pion and
the $\pi$-$\pi$ scattering. On the other hand we also may regard them as effective parameters
for a specific purpose~\cite{oka}, for instance in order to describe the baryons in the HSM. 
In this paper, we will treat $f_{\pi}$ and $e$ as the model parameters with which we can predict 
the physical quantities comparable to the corresponding experimental values 
as shown in Table I. Next the quartic term is necessary to stabilize the 
soliton in the baryon sector. 

Now we investigate baryon phenomenology by using the hyperspherical three metric
\beq
ds^{2}=\lambda^{2}d\mu^{2}+\lambda^{2}\sin^{2}\mu~(d\theta^{2}+\sin^{2}\theta~d\phi^{2}),
\label{metrics30}
\eeq
where the ranges of the hyperspherical coordinates are given by $0\le\mu\le\pi$, $0\le\theta\le\pi$ and $0\le\phi\le 2\pi$, and 
$\lambda$ ($0\le \lambda<\infty$) is the radius parameter of $S^{3}$. Note that $S^{3}$ is described in terms of the 
three dimensional hypersphere coordinates $(\mu,\theta,\phi)$ together with the radius parameter $\lambda$. 
Next, we obtain the soliton energy in the HSM
\beq
E=\frac{f_{\pi}}{e}\left[2\pi L\int_{0}^{\pi}d\mu\sin^{2}\mu\left(\left(\frac{d f}{d\mu}
    +\frac{1}{L}\frac{\sin^{2}f}{\sin^{2}\mu}\right)^{2}+2\left(\frac{1}{L}\frac{d f} 
     {d\mu}+1\right)^{2}\frac{\sin^{2}f}{\sin^{2}\mu}\right)+6\pi^{2}\right],
\label{e2}
\eeq
where $L=ef_{\pi}\lambda$ ($0\le L<\infty$) is a radius parameter 
expressed in dimensionless units. Here $f(\mu)$ is the profile function for 
the hypersphere soliton, and satisfies $f(0)=\pi$ and $f(\pi)=0$ for unit baryon number. Note that one of the advantages of the HSM is that the nonlinear differential equations to be solved numerically in the standard Skyrmion model can 
be linearized and solved analytically in $S^{3}$ hypersphere~\cite{man1,hong98plb}.

In the HSM, spin and isospin states can be treated
by collective coordinates $a^{\mu}=(a^{0},\vec{a})$ $(\mu=0,1,2,3)$, corresponding to the spin and isospin 
collective rotation $A(t)\in$ SU(2) given by $A(t) = a^{0}+i\vec{a}\cdot\vec{\tau}$. Exploiting the coordinates $a^{\mu}$, 
we obtain the Hamiltonian of the form
\begin{equation}
H=E+\frac{1}{8{\cal I}}\pi^{\mu}\pi^{\mu},
\label{hamil61}
\end{equation}
where $\pi^{\mu}$ are canonical momenta conjugate to the collective coordinates $a^{\mu}$. Here the soliton energy lower bound $E$ and 
moment of inertia ${\cal I}$ are given by 
\bea
E&=&\frac{6\pi^{2}f_{\pi}}{e},
\label{deltae}\\
{\cal I}&=&\frac{3\pi^{2}}{e^{3}f_{\pi}}.
\label{calipion}
\eea

Note that, in order to obtain $E$ in (\ref{deltae}) which is now the BPS topological 
lower bound in the soliton energy~\cite{faddeev,man1,manton2,hong98plb}, in  (\ref{e2}) we have used the identity map 
$f(\mu)=\pi-\mu$ with the fixed value $L=L_{B}$ where 
\beq
L_{B}\equiv ef_{\pi}\lambda_{B}=1.\label{lbequiv} 
\eeq
In addition, the identity map with condition $L=L_{B}$ is used to predict the physical quantities such as the moment of inertia ${\cal I}$
in  (\ref{calipion}), baryon masses, charge radii, magnetic moments, axial coupling constant $g_{A}$ and intrinsic pulsation 
frequencies $\omega_{I}$ in the HSM. Note also that the hypersphere coordinates $(\mu,\theta,\phi)$ are integrated out in  (\ref{e2}), and $E$ in 
(\ref{deltae}) is a function of $\lambda_{B}=\frac{1}{ef_{\pi}}$ or equivalently $f_{\pi}$ and $e$ only. Note that the integral expressions of 
$\langle r^{2}\rangle_{E,I=1}$ and 
$g_{A}$ are given by (\ref{r2exp}) and (\ref{gacoupling}) below, respectively, and $\omega_{I}$ is trivially obtainable in terms of ${\cal I}$. Moreover, the above physical quantities except $\langle r^{2}\rangle_{E,I=1}$, $g_{A}$ and $\omega_{I}$ are also given by integral expressions similar to $E$ in (\ref{e2})~\cite{hong98plb}. 
As a result, after integrating out the hypersphere coordinates $(\mu,\theta,\phi)$, these physical quantities are formulated in terms of 
$f_{\pi}$ and $e$ only, as 
shown in  (\ref{calipion}) and  (\ref{mn})--(\ref{radii4}), (\ref{mn2})--(\ref{gacoupling}) and (\ref{omegatheo}).

After performing the canonical quantization in the second class formalism in the HSM, we now construct the Hamiltonian spectrum
\beq
\langle H\rangle=E+\frac{1}{2{\cal I}}I(I+1),
\label{canh0}
\eeq
where $I$ $(=1/2,~3/2,...)$ are baryon isospin quantum numbers. Exploiting  (\ref{canh0}) 
we find the nucleon mass $M_{N}$ for $I=1/2$ and delta baryon mass $M_{\Delta}$ for $I=3/2$, respectively
\beq
M_{N}=ef_{\pi}\left(\frac{6\pi^{2}}{e^{2}}+\frac{e^{2}}{8\pi^{2}}\right),~~~
M_{\Delta}=ef_{\pi}\left(\frac{6\pi^{2}}{e^{2}}+\frac{5e^{2}}{8\pi^{2}}\right).
\label{mn}
\eeq

Next, in the HSM we obtain the magnetic isovector mean square charge radius~\cite{hong98plb}
\beq
\langle r^{2}\rangle_{M,I=1}=\frac{2 ef_{\pi}}{3{\tilde{\cal I}}}\int_{S^{3}}dA_{3}\sin^{2}\mu\sin^{2}f
\left(1+\left(\frac{d f}{d\mu}\right)^{2}+\frac{\sin^{2}f}{\sin^{2}\mu}\right)=\frac{5}{6e^{2}f_{\pi}^{2}},
\label{r2exp}
\eeq
where $\tilde{{\cal I}}$ is the dimensionless moment of inertia defined as $\tilde{{\cal I}}=e^{3}f_{\pi}{\cal I}$ and 
$dA_{3}=\lambda_{B}^{3}\sin^{2}\mu\sin\theta~d\mu~d\theta~d\phi$. Similarly, we find the other charge radii in the HSM 
to yield~\cite{hong98plb}
\bea
\langle r^{2}\rangle^{1/2}_{M,I=0}&=&\langle r^{2}\rangle^{1/2}_{M,I=1}
=\langle r^{2}\rangle^{1/2}_{M,p}=\langle r^{2}\rangle^{1/2}_{M,n}
=\langle r^{2}\rangle^{1/2}_{E,I=1}=\sqrt{\frac{5}{6}}\frac{1}{ef_{\pi}},\label{radii40}\\
\langle r^{2}\rangle^{1/2}_{E,I=0}&=&\frac{\sqrt{3}}{2}\frac{1}{ef_{\pi}},~~~
\langle r^{2}\rangle_{p}=\frac{19}{24}\frac{1}{(ef_{\pi})^{2}},~~~
\langle r^{2}\rangle_{n}=-\frac{1}{24}\frac{1}{(ef_{\pi})^{2}},\label{radii4}
\eea
where the subscripts $E$ and $M$ denote the electric and magnetic radii, respectively. 
Using the charge radii in  (\ref{radii40}) we choose $\langle r^{2}\rangle_{M,p}^{1/2}=0.80$ fm as an 
input parameter. One can then have 
\beq
ef_{\pi}=225.23~{\rm MeV}=(0.876~{\rm fm})^{-1}\label{efpi}
\eeq 
and, with this fixed value of $ef_{\pi}$, one can proceed to calculate the other charge radii as shown in Prediction I in Table~\ref{tablestatic}.\footnote{For 
the experimental data except $\langle r^{2}\rangle_{n}$ we refer to Ref.~\cite{hong98plb}. 
The experimental value of $\langle r^{2}\rangle_{n}$ is given by $-(0.341~{\rm fm})^{2}$~\cite{kopechy}.} 

\section{Baryon physical properties in the first class Dirac quantization in HSM}
\setcounter{equation}{0}
\renewcommand{\theequation}{\arabic{section}.\arabic{equation}}

The model predictions of phenomenological soliton have been shown to include rigorous treatments of geometrical 
constraints because these constraints affect the predictions themselves~\cite{hong15}. Motivated by this, 
we consider the first class Dirac formalism~\cite{di,hong15}. Before we study the first class formalism, we digress 
to note that, in the hypersphere soliton, we have the second class constraints 
\beq
\Omega_{1} = a^{\mu}a^{\mu}-1\approx 0,~~~\Omega_{2} = a^{\mu}\pi^{\mu}\approx 0,
\eeq
so that we have the Poisson algebra 
$\Delta_{k k^{\prime}}=\{\Omega_{k},\Omega_{k^{\prime}}\} = 2\epsilon^{k
k^{\prime}}a^{\mu}a^{\mu}$. The HSM thus becomes a second class constrained system. 

Note that the Dirac quantization should be applied to the hypersphere soliton, in order to predict 
more rigorously physical quantities by using the first class formalism. Following the Dirac Hamiltonian scheme, we find 
the first class constraints
\beq
\tilde{\Omega}_{1}=a^{\mu}a^{\mu}-1+2\theta=0,~~~
\tilde{\Omega}_{2}=a^{\mu}\pi^{\mu}-a^{\mu}a^{\mu}\pi_{\theta}=0,
\label{constfirst}
\eeq
where $(\theta,\pi_{\theta})$ are the St\"uckelberg fields. The first class constraints now satisfy 
$\{\tilde{\Omega}_{1},\tilde{\Omega}_{2}\}=0$. We next obtain the first class Hamiltonian of the form
\beq
\tilde{H}=E+\frac{1}{8{\cal I}}(\pi^{\mu}-a^{\mu}\pi_{\theta})(\pi^{\mu}-a^{\mu}\pi_{\theta})
\frac{a^{\nu}a^{\nu}}{a^{\nu}a^{\nu}+2\theta}.
\label{hct61}
\end{equation}
Note that the first class Hamiltonian is strongly involutive with the first class constraints 
$\{\tilde{\Omega}_{i},\tilde{H}\}=0$ $(i=1, 2)$. After performing the Dirac quantization, we obtain the 
Hamiltonian spectrum with WOC~\cite{hong15} 
\beq
\langle \tilde{H}\rangle=E+\frac{1}{2{\cal I}}\left[I(I+1)+\frac{1}{4}\right].
\label{canh}
\eeq
Here, comparing with the canonical quantization spectrum result $\langle H\rangle$ in  (\ref{canh0}), $\langle \tilde{H}\rangle$ 
obtained via the Dirac quantization with WOC has the additional term $\frac{1}{8{\cal I}}$ in  (\ref{canh}). This additional 
contribution originates from the first class constraints in  (\ref{constfirst}). Using  (\ref{canh}) we newly obtain 
the nucleon and delta baryon masses, respectively
\beq
M_{N}=ef_{\pi}\left(\frac{6\pi^{2}}{e^{2}}+\frac{e^{2}}{6\pi^{2}}\right),~~~
M_{\Delta}=ef_{\pi}\left(\frac{6\pi^{2}}{e^{2}}+\frac{2e^{2}}{3\pi^{2}}\right).\label{mn2}
\eeq
Next we find the magnetic moments~\cite{hong98plb} 
\bea
\mu_{p}&=&\frac{2M_{N}}{ef_{\pi}}\left(\frac{e^{2}}{48\pi^{2}}+\frac{\pi^{2}}{2e^{2}}\right),~~~~~~~
\mu_{n}=\frac{2M_{N}}{ef_{\pi}}\left(\frac{e^{2}}{48\pi^{2}}-\frac{\pi^{2}}{2e^{2}}\right),\nn\\
\mu_{\Delta^{++}}&=&\frac{2M_{N}}{ef_{\pi}}\left(\frac{e^{2}}{16\pi^{2}}+\frac{9\pi^{2}}{10e^{2}}\right),~~~
\mu_{N\Delta}=\frac{2M_{N}}{ef_{\pi}}\cdot\frac{\sqrt{2}\pi^{2}}{2e^{2}},
\label{mus}
\eea
where $M_{N}$ is now given by the nucleon mass with WOC in (\ref{mn2}).

\begin{table}[t]
\caption{The baryon physical properties predicted in the HSM and standard Skyrmion model, compared with experimental data.
In Predictions I and II, the model parameters ($f_{\pi}=58$ MeV, $e=3.89$) and ($f_{\pi}=64$ MeV, $e=4.49$) are exploited in 
the first class Dirac quantization with WOC, respectively. In Prediction III, the model parameters ($f_{\pi}=65$ MeV, $e=5.45$) are used 
in the standard Skyrmion model. The input parameters are indicated by $*$.}
\begin{center}
\begin{tabular}{crrrr}
\hline
Quantity   &Prediction I &Prediction II &Prediction III &Experiment\\
\hline
$\langle r^{2}\rangle^{1/2}_{M,I=0}$ &0.80 {\rm fm} &0.63 {\rm fm} &0.92 {\rm fm} &0.81 {\rm fm}\\ 
$\langle r^{2}\rangle^{1/2}_{M,I=1}$ &0.80 {\rm fm} &0.63 {\rm fm} &$\infty$ &0.80 {\rm fm}\\ 
$\langle r^{2}\rangle^{1/2}_{M,p}$  &0.80 {\rm fm$^{*}$} &0.63 {\rm fm} &$\infty$ &0.80 {\rm fm}\\
$\langle r^{2}\rangle^{1/2}_{M,n}$  &0.80 {\rm fm}  &0.63 {\rm fm} &$\infty$ &0.79 {\rm fm}\\
$\langle r^{2}\rangle^{1/2}_{E,I=0}$  &0.76 {\rm fm} &0.60 {\rm fm} &0.59 {\rm fm} &0.72 {\rm fm}\\ 
$\langle r^{2}\rangle^{1/2}_{E,I=1}$  &0.80 {\rm fm} &0.63 {\rm fm} &$\infty$  &0.88 {\rm fm}\\
$\langle r^{2}\rangle_{p}$  &(0.780 {\rm fm})$^{2}$ &(0.61 {\rm fm})$^{2}$ &$\infty$  &(0.805 {\rm fm})$^{2}$\\
$\langle r^{2}\rangle_{n}$  &$-(0.179~{\rm fm})^{2}$ &$-(0.14~{\rm fm})^{2}$ &$-\infty$ &$-(0.341~{\rm fm})^{2}$\\
$\mu_{p}$  &2.98 &1.88 &1.87  &2.79\\
$\mu_{n}$  &$-2.45$ &$-1.32$ &$-1.31$ &$-1.91$\\
$\mu_{\Delta^{++}}$  &5.69  &3.72  &3.72 &$4.7-6.7$\\
$\mu_{N\Delta}$  &3.84 &2.27 &2.27 &3.29\\
$M_{N}$ &939 {\rm MeV$^{*}$} &939 {\rm MeV$^{*}$} &939 {\rm MeV$^{*}$} &939 {\rm MeV}\\
$M_{\Delta}$ &1112 {\rm MeV} &1232 {\rm MeV$^{*}$} &1232 {\rm MeV$^{*}$} &1232 {\rm MeV}\\
$g_{A}$ &1.30 &0.98 &0.61 &$1.23$\\
\hline
\end{tabular}
\end{center}
\label{tablestatic}
\end{table}

Now, in Prediction I, we fix the model parameters ($f_{\pi}=58$ MeV, $e=3.89$), to reproduce the experimental values for 
$M_{N}=939$ MeV and $\langle r^{2}\rangle_{M,p}^{1/2}=0.80$ fm.\footnote{Note that, in the previous results~\cite{hong98plb} obtained in the second class canonical quantization in the HSM, we have used the model parameters $f_{\pi}=56$ MeV and $e=4.03$ so that the physical quantities 
$M_{N}$, $M_{\Delta}$, $\mu_{p}$, $\mu_{n}$, $\mu_{\Delta^{++}}$ and $\mu_{N\Delta}$ can fit the corresponding experimental data as 
well as possible. In this case we have the prediction $M_{N}=868~{\rm MeV}$ for instance.} Next, exploiting the above 
values of $f_{\pi}$ and $e$, we evaluate the delta baryon mass to yield $M_{\Delta}=1112~{\rm MeV}$. This predicted value for delta baryon mass with WOC is 
improved comparing with the value $M_{\Delta}=1097~{\rm MeV}$ obtained by using the canonical quantization formula in (\ref{mn}). 
Next, in the HSM we newly obtain the axial coupling constant
\beq
g_{A}=\frac{4\pi}{e^{2}}\int_{0}^{\pi}d\mu\sin^{2}\mu\left(1+\cos\mu\right)=\frac{2\pi^{2}}{e^{2}},\label{gacoupling}
\eeq
where we also have used the identity map $f(\mu)=\pi-\mu$ with the fixed value $L=L_{B}$. The theoretical prediction of $g_{A}$ is given by $1.30$, which is comparable to its experimental value $1.23$~\cite{anw83}.
The predictions for the physical quantities are listed in Prediction I. Note that the predicted values for $\mu_{\Delta^{++}}$, $\langle r^{2}\rangle^{1/2}_{M,I=0}$, 
$\langle r^{2}\rangle^{1/2}_{M,I=1}$,  $\langle r^{2}\rangle^{1/2}_{M,n}$ (in addition to the input parameters $M_{N}$ and 
$\langle r^{2}\rangle^{1/2}_{M,p}$) are almost the same as the corresponding experimental data. Next the predictions 
for $g_{A}$, $\langle r^{2}\rangle^{1/2}_{E,I=0}$ and $\langle r^{2}\rangle_{p}$ are within about 6 \% of the experimental values, while 
those for $M_{\Delta}$, $\mu_{p}$ and  $\langle r^{2}\rangle^{1/2}_{E,I=1}$ are within about 10 \% of the experimental ones.  

Next, in Prediction II, we predict the baryon physical quantities using the model parameters 
($f_{\pi}=64$ MeV, $e=4.49$) in the Dirac quantization with WOC. These predictions except the axial coupling constant and charge radii are 
almost same as those in Prediction III~\cite{anw83,hong98plb,adkins} evaluated in the standard Skyrmion model which uses the model parameters 
($f_{\pi}=65$ MeV, $e=5.45$). Note that both in Prediction II and in Prediction III, we exploit the same input 
parameters $M_{N}=939$ MeV and $M_{\Delta}=1232$ MeV to compare these two predictions effectively. The prediction of the axial coupling constant is improved in Prediction II, compared with that in Prediction III. Note that the charge radii in Predictions I and II have finite values comparable to the corresponding experimental data. In contrast, some predicted values for the charge radii in Prediction III are infinite ones.

Now, it seems appropriate to comment on the hypersurface area $A_{3}$ of the hypersphere $S^{3}$ of radius parameter 
$\lambda_{B}$, and the charge radius $\langle r^{2}\rangle_{M,I=1}^{1/2}$ in (\ref{r2exp}). Exploiting the hyperspherical three metric in (\ref{metrics30}), we find that $A_{3}$ can be analyzed in terms of three arc length elements $\lambda_{B}d\mu$, $\lambda_{B}\sin\mu d\theta$ and $\lambda_{B}\sin\mu\sin\theta d\phi$, from which we readily find the three dimensional hypersurface area $A_{3}=2\pi^{2}\lambda_{B}^{3}$. Here note that, since on the thin bubble-like $S^{3}$ soliton  there exists no line element such as $dr$ defined in spherical coordinates $(r,\theta,\phi)$, we do not have a volume and instead we obtain the three dimensional area $A_{3}$ for the hypersurface $S^{3}$. We can then define $\lambda_{B}$ as the radial distance from the center of $S^{3}$ to the hypersurface $S^{3}$ in $R^{4}$. In fact, inserting the value 
$ef_{\pi}=(0.876~{\rm fm})^{-1}$ in  (\ref{efpi}) into the condition $L_{B}=1$ in  (\ref{lbequiv}), in our HSM  we obtain the fixed radius parameter given by $\lambda_{B}=\frac{1}{ef_{\pi}}=0.876~{\rm fm}$. On the other hand, the charge radius $\langle r^{2}\rangle_{M,I=1}^{1/2}$ is the {\it physical quantity} expressed in (\ref{r2exp}). Integrating over a relevant surface density on $S^{3}$ corresponding to the integrand in (\ref{r2exp}), we evaluate $\langle r^{2}\rangle_{M,I=1}$ which is now independent of $\mu$, to yield a specific value of the magnetic isovector root {\it mean} square charge radius. The {\it calculated charge radius} then can be defined as the fixed radial distance to the point on a hypersurface manifold which does not need to be located only on the compact manifold $S^{3}$ of radius parameter $\lambda_{B}$. This hypersurface manifold is now a submanifold in $R^{4}$ which is located at $r=0.80$ fm far from the center of $S^{3}$.
Note that $\langle r^{2}\rangle_{M,I=1}^{1/2}$ denotes the radial distance which is a geometrical invariant giving the same value both in $R^{3}$ (for instance in volume $R^{3}$ which contains the center of $S^{3}$ and is described in terms of $(r,\theta,\phi)$ at $\mu=\frac{\pi}{2}$) and in $R^{4}$. Next, the physical quantity $\langle r^{2}\rangle_{M,I=1}^{1/2}$ calculated in $R^{3}$ (and in $R^{4}$) then can be compared with the corresponding experimental value, similar to the other physical quantities such as $M_{N}$ and $M_{\Delta}$.\footnote{As a toy model of soliton embedded in $R^{3}$, we consider a uniformly charged manifold 
$S^{2}$ described in terms of $(\theta, \phi)$ and a fixed radius parameter $\lambda_{B}$ 
where we have $A_{2}=4\pi\lambda_{B}^{2}$. By integrating over a surface charge density residing on 
$S^{2}$, one can calculate the {\it physical quantity} such as the electric potential, at an arbitrary observation point which does not 
need to be located only on the compact manifold $S^{2}$ of radius parameter $\lambda_{B}$. Next, since the $S^{2}$ soliton of fixed radius parameter $\lambda_{B}$ is {\it embedded} in $R^{3}$, we manifestly define an arbitrary radial distance from the center of the 
compact manifold to an observation point which is located in $R^{3}=S^{2}\times R$. Here $S^{2}$ denotes foliation leaves~\cite{foli} of spherical shell of radius parameter $\lambda$ ($0\le\lambda<\infty$) and $R$ is a manifold associated with radial distance.  Note that the radial distance itself is a fixed geometrical invariant producing the same value both in $R^{2}$ (for instance on equatorial 
plane $R^{2}$ which contains the center of $S^{2}$ and is delineated by $(r, \phi)$ at 
$\theta=\frac{\pi}{2}$) and in $R^{3}$. The same mathematical logic can be applied to $S^{3}$ soliton of fixed radius parameter 
$\lambda_{B}$ {\it embedded} in $R^{4}=S^{3}\times R$ where $S^{3}$ stands for foliation leaves of hyperspherical shell of radius parameter $\lambda$ ($0\le\lambda<\infty$) and $R$ is a manifold related with radial distance.}

Finally it seems appropriate to comment on the Betti numbers associated with the manifold $S^{3}$ in the HSM. First of all, the $p$-th Betti number 
$b_{p}(M)$ is defined as the maximal number of $p$-cycles on $M$:
\beq
b_{p}(M)={\rm dim}~H_{p}(M),
\eeq
where $H_{p}(M)$ is the homology group of the manifold $M$~\cite{derham,derham2,derham3}. For the case of $S^{3}$, we obtain
\bea
H_{0}(S^{3})&=&H_{3}(S^{3})={\mathbf Z},\nn\\
H_{p}(S^{3})&=&0,~{\rm otherwise}.
\eea
The non-vanishing Betti numbers related with $S^{3}$ are thus given by $b_{0}(S^{3})=b_{3}(S^{3})=1$.

\section{Intrinsic frequencies of pulsating baryons}
\setcounter{equation}{0}
\renewcommand{\theequation}{\arabic{section}.\arabic{equation}}
\label{baryonintrinsic}

In this section, we investigate the baryon intrinsic frequencies in the HSM. To do this, we 
introduce an additional term proportional to the first class constraint $\tilde{\Omega}_{2}$ 
into $\tilde{H}$ in  (\ref{hct61}) to obtain the equivalent first class
Hamiltonian
\begin{equation}
\tilde{H}^{\prime}=\tilde{H}+\frac{1}{4{\cal I}}\pi_{\theta}\tilde{\Omega}_{2}, \label{hctpsu2}
\end{equation}
which naturally generates the Gauss law constraints 
\beq
\{\tilde{\Omega}_{1},\tilde{H}^{\prime}\}=\frac{1}{2{\cal I}}\tilde{\Omega}_{2},~~~
\{\tilde{\Omega}_{2},\tilde{H}^{\prime}\}=0. 
\label{gausslaw}
\eeq
Exploiting the Hamiltonian in  (\ref{hctpsu2}), we obtain\footnote{Note that $\langle \tilde{H}^{\prime}\rangle$ with 
$\tilde{H}^{\prime}=\tilde{H}+\frac{1}{4{\cal I}}\pi_{\theta}\tilde{\Omega}_{2}$ is the same as 
$\langle \tilde{H}\rangle$ in  (\ref{canh}) since $\tilde{\Omega}_{2}$ term does not affect the Hamiltonian 
spectrum.}
\beq
\langle \tilde{H}^{\prime}\rangle=E+\frac{1}{2{\cal I}}\left[I(I+1)+\frac{1}{4}\right],
\label{canh2}
\eeq
to yield the predicted values for $M_{N}$ and $M_{\Delta}$ which are the same as the previous ones, 
$M_{N}=939~{\rm MeV}$ and $M_{\Delta}=1112~{\rm MeV}$.

For the first class physical variable $\tilde{W}$ 
governed by the Hamiltonian $\tilde{H}^{\pr}$ in  (\ref{hctpsu2}), the equation of motion in the Poisson bracket form 
is given by~\cite{sakurai}
\beq
\dot{\tilde{W}}=\{\tilde{W},\tilde{H}^{\pr}\},
\label{eomcl}
\eeq
where overdot denotes time derivative. Exploiting the equation of motion in  (\ref{eomcl}), we end up with 
\beq
\dot{\tilde{a}}^{\mu}=\{\tilde{a}^{\mu},\tilde{H}^{\pr}\}=\frac{1}{4{\cal I}}\tilde{\pi}^{\mu},~~~
\dot{\tilde{\pi}}^{\mu}=\{\tilde{\pi}^{\mu},\tilde{H}^{\pr}\}
=-\frac{1}{4{\cal I}}\tilde{\pi}^{\nu}\tilde{\pi}^{\nu}\tilde{a}^{\mu}.
\eeq
Here the first class physical fields $\tilde{a}^{\mu}$ and $\tilde{\pi}^{\mu}$ 
are given as follows~\cite{hong15}
\beq
\tilde{a}^{\mu}=a^{\mu}\left(\frac{a^{\nu}a^{\nu}+2\theta}{a^{\nu}a^{\nu}}\right)^{1/2},~~~
\tilde{\pi}^{\mu}=(\pi^{\mu}-a^{\mu}\pi_{\theta})\left(\frac{a^{\nu}a^{\nu}}{a^{\nu}a^{\nu}+2\theta}\right)^{1/2},
\eeq
which fulfill $\{\tilde{\Omega}_{i},\tilde{a}^{\mu}\}=\{\tilde{\Omega}_{i},\tilde{\pi}^{\mu}\}=0$ ($i=1,2$). 
Here note that we can recover the first class constraint $\tilde{a}^{\mu}\tilde{a}^{\mu}=1$. 
Next we have the identities among the physical fields
\bea
\{\tilde{a}^{\mu},\tilde{\pi}^{\nu}\}&=&\delta^{\mu\nu}-\tilde{a}^{\mu}\tilde{a}^{\nu},~~~~~~~~~~~
\{\tilde{\pi}^{\mu},\tilde{\pi}^{\nu}\}=\tilde{\pi}^{\mu}\tilde{a}^{\nu}-\tilde{a}^{\mu}\tilde{\pi}^{\nu},\nn\\
\{\tilde{a}^{\mu},\tilde{H}\}&=&\frac{1}{4{\cal I}}(\tilde{\pi}^{\mu}
-\tilde{a}^{\mu}\tilde{a}^{\nu}\tilde{\pi}^{\nu}),~~~
\{\tilde{\pi}^{\mu},\tilde{H}\}=\frac{1}{4{\cal I}}
(\tilde{\pi}^{\mu}\tilde{a}^{\nu}\tilde{\pi}^{\nu}-\tilde{a}^{\mu}\tilde{\pi}^{\nu}\tilde{\pi}^{\nu}),\nn\\
\{\tilde{a}^{\mu},\pi_{\theta}\}&=&\tilde{a}^{\mu},~~~~~~~~~~~~~~~~~~~~~~\{\tilde{\pi}^{\mu},\pi_{\theta}\}=-\tilde{\pi}^{\mu}.
\label{id4} 
\eea
Moreover we find 
\beq
\ddot{\tilde{a}}^{\mu}=\{\dot{\tilde{a}}^{\mu},\tilde{H}^{\pr}\}=\frac{1}{4{\cal I}}\dot{\tilde{\pi}}^{\mu}.
\label{ddota}
\eeq

Exploiting  (\ref{canh2}) and (\ref{ddota}), we proceed to obtain
\beq
\ddot{\tilde{a}}^{\mu}=-\frac{1}{4{\cal I}^{2}}\left[I(I+1)+\frac{1}{4}\right]\tilde{a}^{\mu},
\label{eomii}
\eeq
where $I$ again denotes the isospin quantum number of the baryon. Note that the equations of motion for $\tilde{a}^{\mu}$ 
in (\ref{eomii}) represent those for harmonic oscillators of the form
\beq
\ddot{\tilde{a}}^{\mu}=-\omega_{I}^{2}\tilde{a}^{\mu},
\label{eomharmonic}
\eeq 
where the intrinsic pulsating frequency of the baryon with the isospin quantum number $I$ is given by
\beq
\omega_{I}=\frac{1}{2{\cal I}}\left[I(I+1)+\frac{1}{4}\right]^{1/2}.
\label{omegatheo}
\eeq
One of the simple solutions for the equation of motion (\ref{eomharmonic}) is given as follows
\beq
\tilde{a}^{\mu}=\frac{1}{\sqrt{2}}\sin\left(\omega_{I}t+\frac{1}{2}\mu\pi\right),~~~(\mu=0,1,2,3),
\eeq
which satisfy the first class constraint $\tilde{a}^{\mu}\tilde{a}^{\mu}=1$.

It seems appropriate to comment on the Gauss law constraints in (\ref{gausslaw}). 
To derive the equations of motion for the harmonic oscillators in  (\ref{eomharmonic}), we 
have used  (\ref{hctpsu2}) and (\ref{eomcl}), where $\tilde{H}^{\prime}$ is exploited instead of $\tilde{H}$. Note that 
$\tilde{H}^{\prime}$ satisfies the Gauss law constraints in  (\ref{gausslaw}). The Gauss law constraints are now physically 
meaningful since using $\tilde{H}^{\prime}$ associated with these constraints is 
crucial to predict $\omega_{I}$ in  (\ref{omegatheo}).

Inserting the model parameters $f_{\pi}$=58 MeV and $e=3.89$ into the formula for ${\cal I}$ in  
(\ref{calipion}), we obtain
\beq
{\cal I}=(115.11~{\rm MeV})^{-1}.
\label{ecalipred}
\eeq
Now exploiting the value for ${\cal I}$ in  (\ref{ecalipred}) and the formula 
for $\omega_{I}$ in  (\ref{omegatheo}), we arrive at
\beq
\omega_{N}=0.87\times 10^{23}~{\rm sec}^{-1},~~~
\omega_{\Delta}=1.74\times 10^{23}~{\rm sec}^{-1}.
\label{omegandelta}
\eeq
Here we observe that the intrinsic pulsating frequency for more massive particle is greater than that for the less massive one, 
namely $\omega_{\Delta}=2 \omega_{N}$.

\section{Conclusions}
\setcounter{equation}{0}
\renewcommand{\theequation}{\arabic{section}.\arabic{equation}}
\label{conclusion}

In summary, we have evaluated the intrinsic pulsating frequencies of the baryons. To do this, we have exploited 
the HSM, where we have constructed the first class Hamiltonian to quantize the hypersphere soliton. 
Next, we have evaluated the baryon physical quantities such as baryon mases, magnetic moments, charge radii and axial 
coupling constant, in Predictions I and II in Table I. These predictions are much better than those in Prediction III which exploits the standard 
Skyrmion model. In particular, the charge radii in Prediction I are in good agreement 
with the corresponding experimental data. Finally, the intrinsic frequency for more massive particle has been shown to be 
greater than that for the less massive one.

\acknowledgments{The author would like to thank the anonymous referee for helpful comments. He was supported by Basic Science Research Program through
the National Research Foundation of Korea funded by the Ministry of Education, NRF-2019R1I1A1A01058449.}


\end{document}